\documentclass[12pt]{article}
\usepackage{amssymb,amsmath,epsfig}

\numberwithin{equation}{section}

\begin{document}

\title{\textbf{Quantum Corrections for ABGB Black Hole}}

\author{M. Sharif \thanks{msharif.math@pu.edu.pk} and Wajiha Javed \\
Department of Mathematics, University of the Punjab,\\
Quaid-e-Azam Campus, Lahore-54590, Pakistan.}

\date{}
\maketitle

\begin{abstract}
In this paper, we study quantum corrections to the temperature and
entropy of a regular Ay\'{o}n-Beato-Garc\'{\i}a-Bronnikov black hole
solution by using tunneling approach beyond semiclassical
approximation. We use the first law of black hole thermodynamics as
a differential of entropy with two parameters, mass and charge. It
is found that the leading order correction to the entropy is of
logarithmic form. In the absence of the charge, i.e., $e=0$, these
corrections approximate the corresponding corrections for the
Schwarzschild black hole.
\end{abstract}

\section{Introduction}

General Relativity describes that black hole (BH) absorbs all the
light that hits the horizon, reflecting nothing, just like a perfect
black body in thermodynamics. Hawking (1974) suggested that BH like
a black body with a finite temperature, emits radiation from the
event horizon by using quantum field theory in curved spacetime,
named as Hawking radiation. Several attempts (Hartle and Hawking
1976; Gibbons and Hawking 1977) have been made to visualize the
Hawking radiation spectrum by using quantum mechanics of a scalar
particle. However, tunneling (Parikh 2004; Parikh and Wilczek 2000;
Srinivasan and Padmanabhan 1999) provides the best way to visualize
the source of radiation. The essential idea of the tunneling
mechanism is that a particle-antiparticle pair is formed close to
the horizon inside a BH. According to this phenomenon, in the
presence of electric field, particles have the ability to penetrate
energy barriers by following trajectories (not allowed classically).

When a particle with positive energy crosses the horizon, it
appears as Hawking radiation. When a particle with negative energy
tunnels inwards, it is absorbed by the BH, hence its mass
decreases and ultimately vanishes. Similarly, motion of the
particle may be in the form of outgoing or ingoing radial null
geodesics. For outgoing and ingoing motion, the corresponding
action becomes complex and real respectively, whereas classically
a particle can fall behind the horizon. The emission rate of the
tunneling particle from the BH is associated with the imaginary
part of the action which, in turn, is related to the Boltzmann
factor for the emission at the Hawking temperature.

Cognola et al. (1995) investigated the first quantum correction to
the entropy for an eternal $4D$ extremal Reissner-Nordstr$\ddot{o}$m
(RN) BH by using the conformal transformation techniques. Bytsenko
et al. (1998a) suggested that the Schwarzschild-de Sitter BH could
be generated due to back-reaction of dilaton coupled matter in the
early universe, which is the solution of quantum corrected equations
of motion. Bytsenko et al. (1998b) evaluated the first quantum
correction to the Bekenstein-Hawking entropy by using Chern-Simions
representation of the $3D$ gravity. Bytsenko et al. (2001)
calculated the first quantum correction to the finite temperature
partition function for a self-interacting massless scalar field by
using dimensional regularization zeta-function techniques.

Elizalde et al. (1999) investigated the existence of a quantum
process (anti-evaporation) opposite to the Hawking radiation
(evaporation) as an evidence for supersymmetry. Nojiri and Odintsov
(1999a, 1999b, 2000, 2001) studied quantum properties of $2D$
charged BHs and BTZ BH. They found quantum corrected $2D$ charged BH
solution. Also, they evaluated the quantum corrections to mass,
charge, Hawking temperature and BH entropy. They discussed quantum
corrections to thermodynamics (and geometry) of the
Schwarzschild-(anti) de Sitter BHs by using large $N$ one-loop
anomaly induced effective action for dilaton coupled matter. The
same authors also discussed quantum correction to the entropy of
expanding universe.

There are two modifications of the tunneling approach, namely,
Parikh-Wilczek radial null geodesic method (Parikh 2004; Parikh and
Wilczek 2000) and the Hamilton-Jacobi method (Srinivasan and
Padmanabhan 1999). Recently, based on the Hamilton-Jacobi method,
Banerjee and Majhi (2008) developed a tunneling formalism beyond
semiclassical approximation. They computed quantum corrections to
the Hawking temperature $T=\frac{\kappa_0}{2\pi}$ and
Bekenstein-Hawking entropy $S_{BH}=\frac{A}{4\hbar}$ (Bekenstein
1972). The first law of thermodynamics also holds in the context of
quantum corrections. When quantum effects are considered, the area
law of BH entropy should undergo corrections using loop quantum
gravity, i.e.,
\begin{equation}
S=S_{BH}+\alpha\ln S_{BH}+\cdots.
\end{equation}
Loop quantization reproduces the result of Bekenstein-Hawking
entropy of BH. This formalism has been applied on various BHs
(Banerjee and Modak 2009; Modak 2009; Zhu et al. 2009a) and FRW universe
model (Zhu et al. 2009b).

Banerjee and Modak (2009) gave a simple approach to obtain the
entropy for any stationary BH. Akbar and Saifullah (2010, 2011) studied
quantum corrections to entropy and horizon area for the
Kerr-Newmann, charged rotating BTZ and Einstein-Maxwell
dilaton-axion BHs. Recently, Larra$\tilde{n}$aga (2011a, 2011b) extended
this work for a charged BH of string theory and for the Kerr-Sen BH.
Majhi (2009) with his collaborator Samanta (2010) analyzed the Hawking radiation
as tunneling of a Dirac particle, photon and a gravitino through an
event horizon by applying the Hamilton-Jacobi method beyond the
semiclassical approximation.

Chen \emph{et al.} (2011) investigated the corrected Hawking
temperature and entropy for various BHs, FRW universe model and
neutral black rings. Jamil and Darabi (2011) studied quantum
corrections to the Hawking temperature, entropy and
Bekenstein-Hawking entropy-area relation for a Braneworld BH by
using tunneling approach beyond semiclassical approximation. In a
recent paper, we have explored these quantum corrections for a
Bardeen regular BH (Sharif and Javed 2010). Also, we have discussed
thermodynamics of Bardeen BH in noncommutative space (Sharif and
Javed 2011).

This paper investigates temperature and entropy corrections for the
Ay\'{o}n-Beato-Garc\'{\i}a-Bronnikov (ABGB) BH which is a
generalization of the entropy correction for the Schwarzschild BH
(Banerjee and Majhi 2008). The motivation to study ABGB black hole
is its feature to express the location of the horizons in terms of
Lambert function which is used in the discussion of the extremal
configurations. Outside the event horizon, this BH solution closely
resembles with the RN geometry both in its local as well as global
structure. Matyjasek (2008) described this BH solution (exists as a
perturbative solution and its various characteristics acquire the
correction) by using quadratic gravity equations.

Here we skip the details of the formulation as it is given in many
papers, e.g. (Sharif and Javed 2010) which consists of basic
material used to evaluate corrections to the entropy and
temperature. The plan of the paper is as follows; In Section
\textbf{2}, we evaluate semiclassical thermodynamical quantities
(temperature and entropy) for the ABGB regular BH. Section
\textbf{3} provides the corrections to these quantities. Finally, in
the last section, we summarize the results.

\section{Thermodynamical Quantities}

When particles with positive energy tunnel across the horizon, a
BH loses its mass. The tunneling amplitude of particles emitted by
a BH in the form of Hawking radiation can be calculated for a
charged regular BH. The general line element of a spherically
symmetric BH is given by
\begin{equation}
{ds}^2=-F{dt}^2+F^{-1}{dr}^2+r^2{d\theta}^2+r^2\sin^2\theta{d\phi}^2,
\label{14}
\end{equation}
where $F=1-2\frac{M(r)}{r}$. This metric can be reduced to
well-known BHs for the special choice of $M(r)$. Ay\'{o}n-Beato and
Garc\'{\i}a (1999) and Bronnikov (2000) formulated
a solution of the coupled system of equations of non-linear
electrodynamics and gravity representing a class of the BHs. This is
given by
\begin{equation}
M(r)=m\left[1-\tanh\left(\frac{e^2}{2mr}\right)\right],\label{15}
\end{equation}
where $m$ is the mass and $e$ is either electric or magnetic
charge. This solution describes a regular static spherically
symmetric configuration which reduces to the Schwarzschild
solution for $e=0$.

The ABGB regular BH solution has a spherical event horizon at
$F(r_+)=0$ or $r_+=2M$, where $r_+$ is the event horizon. Replacing
the value of $M$, $F(r)$ will take the following form
\begin{equation}
F(r)=1-\frac{2m}{r}\left[1-\tanh\left(\frac{e^2}{2mr}\right)\right],
\label{16}
\end{equation}
whose roots are given in (Matyjasek 2007, 2008) and its area is
(Larra$\tilde{n}$aga 2011a, 2011b)
\begin{equation}
A=\int\sqrt{g_{\theta\theta}g_{\phi\phi}}d\theta d\phi=4\pi
r_+^2.\label{24}
\end{equation}
In terms of power series, the ABGB solution turns out to be
\begin{equation}
F(r)=1-\frac{2m}{r}+\frac{e^2}{r^2}-\frac{e^6}{12m^2r^4}+
\textsl{O}(\frac{1}{r^6}).\label{}
\end{equation}
Notice that $F(r)$ differs from the Reissner-Nordstr\"{o}m (RN)
solution by terms of order $\textsl{O}(e^6)$. For small $e$, we can
neglect terms of order $\textsl{O}(e^6)$ and onward and hence
exactly reduces to the RN solution.

Here we assume that the terms of order $\textsl{O}(\frac{1}{r^6})$
and the higher orders can be neglected. Thus $F(r)$ can be written
as follows
\begin{equation}
F(r)=1-\frac{2m}{r}+\frac{e^2}{r^2}-\frac{e^6}{12m^2r^4}.\label{17}
\end{equation}
From this equation, $F(r)=0$ leads to cubic equation in $m$, i.e.,
\begin{equation}
m^3-\frac{r}{2}(1+\frac{e^2}{r^2})m^2+\frac{e^6}{24r^3}=0.
\end{equation}
Using \emph{Cardan's solution} (Nickalls 1993), we can evaluate the only
real root of this equation, i.e.,
\begin{eqnarray}
m&=&\frac{r_+}{6}(1+\frac{e^2}{r_+^2})+\left[\frac{1}{2}\left(-\frac{e^6}{24r_+^3}+
\frac{2r_+^3}{216}(1+\frac{e^2}{r_+^2})^3\right.\right.\nonumber\\&+&
\left.\left.\sqrt{\frac{e^{12}}{576r_+^6}-\frac{e^6}{1296}
(1+\frac{e^2}{r_+^2})^3}\right)\right]^{\frac{1}{3}}+\left[\frac{1}{2}
\left(-\frac{e^6}{24r_+^3}+\frac{2r_+^3}{216}(1+\frac{e^2}{r_+^2})^3
\right.\right.\nonumber\\
&-&\left.\left.\sqrt{\frac{e^{12}}{576r_+^6}-
\frac{e^6}{1296}(1+\frac{e^2}{r_+^2})^3}\right)\right]^{\frac{1}{3}}.\label{h}
\end{eqnarray}
For $e=0$, this reduces to Schwarzschild case whose horizon radius is
$r_+=2m$.

Now we simplify Eq.(\ref{h}) by Taylor
series upto first order approximation. The term in Eq.(\ref{h})
can be written as
\begin{equation}
\sqrt{\frac{e^{12}}{576r_+^6}-\frac{e^6}{1296}(1+\frac{e^2}{r_+^2})^3}\approx
\frac{\sqrt{5}e^6}{72r_+^3}-\frac{e^4}{12\sqrt{5}r_+}-\frac{e^2r_+}{12\sqrt{5}}-
\frac{r_+^3}{36\sqrt{5}}.
\end{equation}
Consequently, the value of $m$ will become
\begin{eqnarray}
m&\approx&\frac{r_+}{6}(1+\frac{e^2}{r_+^2})+r_+(\frac{1}{216}-
\frac{1}{72\sqrt{5}})^{\frac{1}{3}}
\left[1+\frac{1}{3(\frac{1}{216}-\frac{1}{72\sqrt{5}})}
\left\{\frac{e^6}{r_+^6}\right.\right.\nonumber\\
&\times&\left.\left.(-\frac{7}{432}+ \frac{\sqrt{5}}{144})+
\frac{e^4}{r_+^4}(\frac{1}{72}-\frac{1}{24\sqrt{5}})+
\frac{e^2}{r_+^2}(\frac{1}{72}-\frac{1}{24\sqrt{5}})\right\}\right]
\nonumber\\&+&r_+(\frac{1}{216}+\frac{1}{72\sqrt{5}})^{\frac{1}{3}}
\left[1+\frac{1}{3(\frac{1}{216}+ \frac{1}{72\sqrt{5}})}
\left\{-\frac{e^6}{r_+^6}(\frac{7}{432}+\frac{\sqrt{5}}{144})
\right.\right.\nonumber\\
&+&\left.\left.\frac{e^4}{r_+^4}
(\frac{1}{72}+\frac{1}{24\sqrt{5}})+\frac{e^2}{r_+^2}(\frac{1}{72}+
\frac{1}{24\sqrt{5}})\right\}\right].\label{b}
\end{eqnarray}
Notice that if the term $(\frac{1}{216}-
\frac{1}{72\sqrt{5}})^{\frac{1}{3}}$ is solved by Taylor series
(which is a divergent series) then the ABGB regular BH mass
exactly reduces to the Schwarzschild BH mass $m=0.5r$, otherwise
it can be defined as
\begin{equation}
m=0.3r_++\frac{0.9e^2}{r_+}-\frac{0.8e^6}{r_+^5}+\frac{0.7e^4}{r_+^3}.\label{a}
\end{equation}
For $e=0$, this expression approximately leads to the
Schwarzschild BH mass.

The semiclassical Hawking temperature
$T_H$ (Akbar 2007; Kothawala et al. 2007) is
\begin{equation}
T_H={\frac{\hbar F'(r)}{4\pi}}|_{r=r_+}=\frac{\hbar}{2\pi}
\left(\frac{m}{r_+^2}-\frac{e^2}{r_+^3}+\frac{e^6}{6m^2r_+^5}\right),
\label{18}
\end{equation}
where $F'(r)$ denotes derivative of $F$ with respect to $r$. The
values of $F$ and $m$ are given by Eqs.(\ref{17}) and (\ref{a})
respectively. Substituting this value of $m$ in Eq.(\ref{18}) and
simplifying, it follows that
\begin{eqnarray}
T_H&=&\frac{\hbar}{2\pi}
\left(\frac{0.3}{r_+}-\frac{0.1e^2}{r_+^3}+\frac{0.6999e^4}{r_+^5}+\textsl{O}(\frac{1}{r_+^7})
\right).\label{c}
\end{eqnarray}
The electric potential is given by (Akbar and Siddiqui 2007)
\begin{equation}
\Phi=\frac{\partial m}{\partial e}|_{r=r_+}=-4.8
\frac{e^5}{r_+^5}+2.8\frac{e^3}{r_+^3}+1.8\frac{e}{r_+}.\label{19}
\end{equation}
The semiclassical entropy has the form
\begin{equation}
S_0(m,e)=\int\frac{dm}{T_H}=\frac{2\pi}{\hbar}
\int\frac{dm}{(\frac{0.3}{r_+}-\frac{0.1e^2}{r_+^3}+\frac{0.6999e^4}{r_+^5})}.\label{20}
\end{equation}
To evaluate this integral, we use Eq.(\ref{a}) which yields
\begin{equation}
dm=\left(0.3-\frac{0.9e^2}{r_+^2}+\frac{4e^6}{r_+^6}-\frac{2.1e^4}{r_+^4}\right)dr_+.\label{21}
\end{equation}
Inserting this value in Eq.(\ref{20}), we obtain
\begin{equation}
S_0=\frac{2\pi}{\hbar}\int\left(r_+-\frac{2.6667e^2}{r_+}-\frac{10.3333e^4}{r_+^3}+
\frac{18e^6}{r_+^5}+\textsl{O}
(\frac{1}{r_+^7})\right)dr_+.\label{abgb1}
\end{equation}
Integrating this equation, it follows that
\begin{equation}
S_0=\frac{\pi}{\hbar}\left(-5.3333e^2\ln
r_++r_+^2+\frac{10.3333e^4}{r_+^2}-\frac{9e^6}{r_+^4}+
\textsl{O}(\frac{1}{r_+^6})\right).\label{g}
\end{equation}
It is interesting to mention here that for $e=0$ and $\hbar=1$, we
recover the Bekenstein-Hawking area law, i.e.,
$S_0=\frac{A}{4}$.

\section{Corrections to the Thermodynamical Quantities}

Here we work out the corrected form of Hawking temperature and the
corresponding entropy for the charged regular BH by taking into
account the quantum effects on the thermodynamical quantities.

\subsection{Hawking Temperature Corrections}

The expression for the semiclassical Hawking temperature (\ref{c})
can be written as
\begin{equation}
T_H\approx\frac{\hbar}{2\pi}
\left(\frac{0.3}{r_+}-\frac{0.1e^2}{r_+^3}+\frac{0.7e^4}{r_+^5}\right).\label{d}
\end{equation}
The corrected temperature is given by (Sharif and Javed 2010)
\begin{equation}
T=T_H\left(1-\frac{\beta\hbar}{m^2}\right),\label{22222}
\end{equation}
where $\beta$ is given by
\begin{equation}
\beta=-\frac{1}{360\pi}\left(-N_0-\frac{7}{4}N_{\frac{1}{2}} +13
N_1+\frac{233}{4}N_{\frac{3}{2}}-212 N_2\right),
\end{equation}
$N_s$ refers to the number of spin $s$ fields (Banerjee and Majhi 2008). Inserting
the value of $m$ in Eq.(\ref{22222}), it follows that
\begin{equation}
T=T_H\left[1-\frac{\beta\hbar}{0.09r_+^2}\left\{1-2\left(\frac{3e^2}{r_+^2}-
\frac{2.6667e^6}{r_+^6}+\frac{2.3333e^4}{r_+^4}\right)\right\}\right].
\label{e}
\end{equation}
Using Eq.(\ref{d}) in (\ref{e}), we obtain the quantum correction of
temperature $T$ by neglecting the higher order terms
\begin{equation}
T\approx\frac{\hbar}{2\pi}\left(\frac{0.3}{r}-\frac{0.1e^2}{r^3}+
\frac{0.7e^4}{r^5}\right)-\frac{\beta{\hbar}^2}{0.18\pi r^2}\left(
\frac{0.3}{r}-\frac{1.9e^2}{r^3}-\frac{0.1e^4}{r^5}\right).\label{f}
\end{equation}
For $e=0$, this implies that $T=\frac{0.15\hbar}{\pi r}\left(1-
\frac{11.11\beta\hbar}{r^2}\right)$, which approaches to the
corrected Hawking temperature of the Schwarzschild BH.

\subsection{Entropy Corrections}

Here, we evaluate the quantum corrections to the entropy of the
ABGB charged regular BH. The corrected form of entropy is (Sharif and Javed 2010)
\begin{equation}
S(r,t)=S_0(r,t)\left(1+\sum_{i}\alpha_i\frac{\hbar^i}{m^{2i}}\right).
\label{2}
\end{equation}
In terms of horizon radius, this can be written as
\begin{equation}
S(r,t)=S_0(r,t) \left(
1+\sum_{i}\frac{\alpha_i\hbar^i}{(0.3r_++\frac{0.9e^2}{r_+}-
\frac{0.8e^6}{r_+^5}+
\frac{0.7e^4}{r_+^3})^{2i}} \right), \label{3}
\end{equation}
where the semiclassical entropy can be written from Eq.(\ref{g}) as
\begin{equation}
S_0\approx\frac{\pi}{\hbar}\left(-5.3333e^2\ln
r_++r_+^2+\frac{10.3333e^4}{r_+^2}-\frac{9e^6}{r_+^4}\right).
\end{equation}
The corrected form of the Hawking temperature is
(Sharif and Javed 2010)
\begin{equation}
T=T_H{\left(
1+\sum_{i}\frac{\alpha_i\hbar^i}{(0.3r_++\frac{0.9e^2}{r_+}-\frac{0.8e^6}{r_+^5}+
\frac{0.7e^4}{r_+^3})^{2i}} \right)}^{-1}.\label{i}
\end{equation}
Using the first law of thermodynamics, $dm=TdS+\Phi de$, we can write the
condition for the exact differential as
\begin{equation}
\frac{\partial}{\partial e}\left(\frac{1}{T}\right)=
\frac{\partial}{\partial m}\left(-\frac{\Phi}{T}\right).\label{10}
\end{equation}
Inserting the value of corrected temperature, it follows that
\begin{eqnarray}
&&\frac{\partial}{\partial e}\left(\frac{1}{T_H}\right){\left(
1+\sum_{i}\frac{\alpha_i\hbar^i}{(0.3r_++\frac{0.9e^2}{r_+}-\frac{0.8e^6}{r_+^5}+
\frac{0.7e^4}{r_+^3})^{2i}} \right)}\nonumber\\&=&
\frac{\partial}{\partial m}\left(-\frac{\Phi}{T_H}\right){\left(
1+\sum_{i}\frac{\alpha_i\hbar^i}{(0.3r_++\frac{0.9e^2}{r_+}-\frac{0.8e^6}{r_+^5}+
\frac{0.7e^4}{r_+^3})^{2i}} \right)}\label{}.
\end{eqnarray}

Using the exact differential condition, the entropy in the
integral form is
\begin{equation}
S(m,e)=\int\frac{1}{T}dm-\int\frac{\Phi}{T}de-\int\left(\frac{\partial}{\partial
e}\left(\int\frac{1}{T}dm\right)\right)de.\label{11}
\end{equation}
Substituting the value of corrected temperature, the
corresponding corrected entropy will become
\begin{eqnarray}
S(m,e)&=&\int\frac{1}{T_H}{\left(
1+\sum_{i}\frac{\alpha_i\hbar^i}{(0.3r_++\frac{0.9e^2}{r_+}-\frac{0.8e^6}{r_+^5}+
\frac{0.7e^4}{r_+^3})^{2i}}
\right)}dm\nonumber\\&-&\int\frac{\Phi}{T_H}{\left(
1+\sum_{i}\frac{\alpha_i\hbar^i}{(0.3r_++\frac{0.9e^2}{r_+}-\frac{0.8e^6}{r_+^5}+
\frac{0.7e^4}{r_+^3})^{2i}}
\right)}de\nonumber\\
&-&\int\left(\frac{\partial}{\partial
e}\left(\int\frac{1}{T_H}\left(
1+\sum_{i}\right.\right.\right.\nonumber\\&&\left.\left.\left.\frac{\alpha_i\hbar^i}
{(0.3r_++\frac{0.9e^2}{r_+}-\frac{0.8e^6}{r_+^5}+
\frac{0.7e^4}{r_+^3})^{2i}}
\right)dm\right)\right)de.\label{abc}
\end{eqnarray}
We can simplify these complicated integrals by employing the
exactness criterion (Sharif and Javed 2010). Consequently, this reduces to
\begin{equation}
S(m,e)=\int\frac{1}{T_H}{\left(
1+\sum_{i}\frac{\alpha_i\hbar^i}{(0.3r_++\frac{0.9e^2}{r_+}-\frac{0.8e^6}{r_+^5}+
\frac{0.7e^4}{r_+^3})^{2i}} \right)}dm\label{}
\end{equation}
which can be written in expanded form as
\begin{eqnarray}\label{1}
S(m,e)&=&\int\frac{1}{T_H}dm+\int{\frac{\alpha_1\hbar}{T_H(0.3r_++\frac{0.9e^2}{r_+}-
\frac{0.8e^6}{r_+^5}+\frac{0.7e^4}{r_+^3})^{2}}}dm\nonumber\\&+&
\int{\frac{\alpha_2\hbar^2}{T_H(0.3r_++\frac{0.9e^2}{r_+}-\frac{0.8e^6}{r_+^5}+
\frac{0.7e^4}{r_+^3})^{4}}}dm\nonumber\\
&+&\int{\frac{\alpha_3\hbar^3}{T_H(0.3r_++\frac{0.9e^2}{r_+}-\frac{0.8e^6}{r_+^5}+
\frac{0.7e^4}{r_+^3})^{6}}}dm+....\nonumber\\
&=&I_1+I_2+I_3+I_4+....
\end{eqnarray}
The first integral $I_1$ has been evaluated in Eq.(\ref{g}) and
$I_2, I_3,...$ are corrections due to quantum effects. Thus
\begin{equation}
I_2=2\pi\alpha_1\int\frac{(0.3-\frac{0.9e^2}{r_+^2}+\frac{4e^6}{r_+^6}-
\frac{2.1e^4}{r_+^4})}{(
\frac{0.3}{r_+}-\frac{0.1e^2}{r_+^3}+\frac{0.7e^4}{r_+^5})(0.3r_++
\frac{0.9e^2}{r_+}-\frac{0.8e^6}{r_+^5}+\frac{0.7e^4}{r_+^3})^2}dr_+,
\label{}
\end{equation}
\begin{equation}
I_3=2\pi\alpha_2\hbar\int\frac{(0.3-\frac{0.9e^2}{r_+^2}+\frac{4e^6}{r_+^6}-
\frac{2.1e^4}{r_+^4})}{(
\frac{0.3}{r_+}-\frac{0.1e^2}{r_+^3}+\frac{0.7e^4}{r_+^5})(0.3r_++\frac{0.9e^2}{r_+}-
\frac{0.8e^6}{r_+^5}+\frac{0.7e^4}{r_+^3})^4}dr_+. \label{}
\end{equation}
In general, we can write for $k>3$
\begin{eqnarray}
I_k&=&2\pi\alpha_{k-1}\hbar^{k-2}
\int\left(\frac{(0.3-\frac{0.9e^2}{r_+^2}+\frac{4e^6}{r_+^6}-\frac{2.1e^4}{r_+^4})}{(
\frac{0.3}{r_+}-\frac{0.1e^2}{r_+^3}+\frac{0.7e^4}{r_+^5})}\right.\nonumber\\
&\times&\left.\frac{1}{(0.3r_++
\frac{0.9e^2}{r_+}-\frac{0.8e^6}{r_+^5}+\frac{0.7e^4}{r_+^3})^{2(k-1)}}\right)dr_+.
\label{}
\end{eqnarray}
Replacing all the values in Eq.(\ref{1}), it follows that
\begin{eqnarray}
S(m,e)&=&2\pi\hbar^{-1}\int\frac{(0.3-\frac{0.9e^2}{r_+^2}+\frac{4e^6}{r_+^6}-
\frac{2.1e^4}{r_+^4})}{(
\frac{0.3}{r_+}-\frac{0.1e^2}{r_+^3}+\frac{0.7e^4}{r_+^5})}dr_+\nonumber\\&+&
2\pi\alpha_1\int\frac{(0.3-\frac{0.9e^2}{r_+^2}+\frac{4e^6}{r_+^6}-\frac{2.1e^4}{r_+^4})}{(
\frac{0.3}{r_+}-\frac{0.1e^2}{r_+^3}+\frac{0.7e^4}{r_+^5})(0.3r_++\frac{0.9e^2}{r_+}-
\frac{0.8e^6}{r_+^5}+
\frac{0.7e^4}{r_+^3})^2}dr_+\nonumber\\&+&\sum_{k>2}
2\pi\alpha_{k-1}\hbar^{k-2}
\int\left(\frac{(0.3-\frac{0.9e^2}{r_+^2}+\frac{4e^6}{r_+^6}-\frac{2.1e^4}{r_+^4})}{(
\frac{0.3}{r_+}-\frac{0.1e^2}{r_+^3}+\frac{0.7e^4}{r_+^5})}\right.\nonumber\\&\times&\left.
\frac{1}{(0.3r_++\frac{0.9e^2}{r_+}-
\frac{0.8e^6}{r_+^5}+\frac{0.7e^4}{r_+^3})^{2(k-1)}}\right)dr_+.
\label{123asbc}
\end{eqnarray}
This gives the quantum correction to the entropy for a ABGB
charged BH.

For $e=0$, Eq.(\ref{123asbc}) reduces to
\begin{equation}
S=\frac{A}{4\hbar}+\frac{\pi\alpha_1{\ln
A}}{(0.3)^2}-\frac{4\pi^2\hbar\alpha_2}{(0.3)^4A}+...,\label{efg}
\end{equation}
where $A$ is given by Eq.(\ref{24}). This is approximately similar
to the corrected entropy of the Schwarzschild BH (Banerjee and Majhi 2008). It is
worth mentioning here that the first term of Eq.(\ref{efg}) is the
semiclassical Bekenstein-Hawking area law, i.e.,
$S_{BH}=\frac{A}{4\hbar}$, while the remaining terms are due to
quantum corrections. Thus, $S_{BH}$ is modified by quantum effects.
Integrating Eq.(\ref{123asbc}), it follows that
\begin{eqnarray}
S(m,e)&=&\pi\hbar^{-1}\left(-5.3333e^2\ln
r_++r_+^2+\frac{10.3333e^4}{r_+^2}
\right.\nonumber\\
&-&\left.\frac{9e^6}{r_+^4}\right)+\pi\alpha_1\left(22.22\ln r_++
\frac{96.22e^2}{r_+^2}\right.\nonumber\\
&+&\left.\frac{27.78e^4}{r_+^4}\right)+\frac{2\pi\hbar{\alpha_2}}{(0.3)^4}
\left(-\frac{0.5}{r_+^2}+\frac{3.67e^2}{r_+^4}\right)+....
\label{123456}
\end{eqnarray}
The entropy (\ref{123asbc}) in terms of $A$ is given as follows
\begin{eqnarray}
S(m,e)&=&\frac{\hbar^{-1}}{4}\int\frac{(1-\frac{3e^2}{(\frac{A}{4\pi})}+
\frac{13.3333e^6}{(\frac{A}{4\pi})^3}-\frac{7e^4}{(\frac{A}{4\pi})^2})}
{(1-\frac{0.3333e^2}{(\frac{A}{4\pi})}+\frac{2.3333e^4}{(\frac{A}{4\pi})^2})}dA
\nonumber\\&+&
\frac{\alpha_1\pi}{(0.3)^2}
\int\left(\frac{(1-\frac{3e^2}{(\frac{A}{4\pi})}+
\frac{13.3333e^6}{(\frac{A}{4\pi})^3}-\frac{7e^4}{(\frac{A}{4\pi})^2})}
{(1-\frac{0.3333e^2}{(\frac{A}{4\pi})}+\frac{2.3333e^4}{(\frac{A}{4\pi})^2})
}\right.\nonumber\\&\times&\left.\frac{1}{(1+\frac{3e^2}{(\frac{A}{4\pi})}-
\frac{2.6667e^6}{(\frac{A}{4\pi})^3}+
\frac{2.3333e^4}{(\frac{A}{4\pi})^2})^2}\right)dA\nonumber\\&+&
\sum_{k>2}\frac{2^{2k-4}\hbar^{k-2}\alpha_{k-1}(\pi)^{k-1}}{(0.3)^{2k-2}}
\int\left(\frac{(1-\frac{3e^2}{(\frac{A}{4\pi})}+
\frac{13.3333e^6}{(\frac{A}{4\pi})^3}-\frac{7e^4}{(\frac{A}{4\pi})^2})}
{A^{k-1}(1-\frac{0.3333e^2}{\frac{A}{4\pi}}+\frac{2.3333e^4}{(\frac{A}{4\pi})^2})
}\right.\nonumber\\&\times&
\left.\frac{1}{(1+\frac{3e^2}{(\frac{A}{4\pi})}-
\frac{2.6667e^6}{(\frac{A}{4\pi})^3}+
\frac{2.3333e^4}{(\frac{A}{4\pi})^2})^{2k-2}}\right)dA.\label{erdsa}
\end{eqnarray}
\begin{eqnarray}
S(m,e)&=&\frac{\hbar^{-1}}{4}\int\left(1-\frac{2.66666e^2}{(\frac{A}{4\pi})}-
\frac{10.3334e^4}{(\frac{A}{4\pi})^2}+\frac{18e^6}{(\frac{A}{4\pi})^3}
\right.\nonumber\\&+&\left.\textsl{O}(\frac{1}{A^4})\right)dA+
\frac{\alpha_1\pi}{(0.3)^2}
\int\frac{1}{A}\left(1-\frac{8.6667e^2}{(\frac{A}{4\pi})}+
\frac{5e^4}{(\frac{A}{4\pi})^2}
\right.\nonumber\\
&+&\left.\frac{60.8855e^6}{(\frac{A}{4\pi})^3}+\textsl{O}(\frac{1}{A^4})\right)dA+
\frac{4\alpha_2\pi^2\hbar}{(0.3)^4}
\int\frac{1}{A^2}\left(1-\frac{14.6667e^2}{(\frac{A}{4\pi})}
\right.\nonumber\\&+&\left.
\frac{20.3335e^4}{(\frac{A}{4\pi})^2}+\frac{103.775e^6}{(\frac{A}{4\pi})^3}+
\textsl{O}(\frac{1}{A^4})\right)dA+....\label{erdsa}
\end{eqnarray}
When we take $e=0$, this equation leads to Eq.(\ref{efg}). Solving
Eq.(\ref{erdsa}), we obtain
\begin{eqnarray}
S(m,e)&=&\frac{\hbar^{-1}}{4}\left(A+\frac{1631.78e^4}{A}-\frac{17859.6e^6}{A^2}+
\textsl{O}(\frac{1}{A^3})\right)\nonumber\\&+&\frac{\alpha_1\pi}{(0.3)^2}
\left(\ln
A+\frac{108.909e^2}{A}-\frac{394.785e^4}{A^2}+\textsl{O}(\frac{1}{A^3})\right)
\nonumber\\&+&\frac{4\alpha_2 \pi^2 \hbar}{(0.3)^4}
\left(-\frac{1}{A}+\frac{92.1536e^2}{A^2}+\textsl{O}(\frac{1}{A^3})\right)+....
\label{rtyu}
\end{eqnarray}

\section{Outlook}

The semiclassical entropy and temperature of the BH should be
corrected due to quantum effects. The tunneling formalism beyond
semiclassical approximation is one of the approaches which provides
quantum corrections to these thermodynamical quantities of a BH. In
general, the corrected form of entropy has a logarithmic leading
order term.

The entropy of the BH can be calculated by using various methods.
For instance, Wald's technique (Wald 1993; Iyer and Wald 1994;
Jacobson et al. 1994) is suitable for higher curvature theories
while some techniques (Whitt 1985; Audretsch et al. 1993; Jacobson
et al. 1995) are based on the field redefinition and Visser's (1992,
1993a, 1993b) Euclidean approach.

In this paper, we use quantum tunneling approach beyond
semiclassical approximation to study the quantum corrections of
temperature and entropy for the ABGB charged regular BH. For this
purpose, first of all, we have evaluated the semiclassical
temperature and entropy that reduce to the temperature and entropy
of the Schwarzschild case (Banerjee and Majhi 2008) for $e=0$. The
quantum corrections to the temperature and entropy approximate to
the corrected form of temperature (Banerjee and Majhi 2008) and
entropy (\ref{efg}) of the Schwarzschild case respectively for zero
charge.

It is interesting to mention here that the leading order entropy
correction of the charged regular BH turns out to be a logarithmic
term which is expected due to quantum effects. The other terms
involve ascending powers of the inverse of the area (Banerjee and
Majhi 2008). The Bekenstein-Hawking entropy-area relationship also
reduces to the Schwarzschild when we take zero charge. It is
worthwhile to note that quantum corrections to the thermodynamical
quantities, i.e., temperature and entropy, given by Eqs.(3.5) and
(3.21) respectively, reduce to the classical temperature and entropy
((2.13) and (2.18)) after the correction is disappeared.

We would like to point out here that semiclassical thermodynamical
quantities and their corresponding corrections has been evaluated by
using Taylor's expansion upto the first order approximation. These
approximations are valid only for the specific ratio of $e$ and $r$.
Consequently, quantum corrections of temperature and entropy with
specific ratio of $e$ and $r$ do not represent class of corrections
corresponding to semiclassical values of temperature and entropy.
Hence quantum corrections to the thermodynamical quantities are not
larger than the semiclassical thermodynamical quantities.

Finally, it is mentioned that the entropy of this BH solution has
also been discussed by Matyjasek. However, he used Wald's and
Visser's Euclidean approaches (Matyjasek 2008). In our work, we have
analyzed the issue of quantum corrections by using Hamilton-Jacobi
method beyond the semiclassical approximation.

\vspace{0.25cm}

{\bf Acknowledgment}

\vspace{0.25cm}

We would like to thank the Higher Education Commission, Islamabad,
Pakistan, for its financial support through the {\it Indigenous
Ph.D. 5000 Fellowship Program Batch-IV}.


\section*{References}

Akbar, M.: Chin. Phys. Lett. {\bf 24}, 1158(2007)\\
Akbar, M., Saifullah, K.: Eur. Phys. J. C {\bf 67}, 205(2010)\\
Akbar, M., Saifullah, K.: Gen. Relativ. Gravit. {\bf 43}, 933(2011)\\
Akbar, M., Siddiqui, A.A.: Phys. Lett. B {\bf 656}, 217(2007)\\
Audretsch, J., et al.: Phys. Rev. D {\bf 47}, 3303(1993)\\
Ayon-Beato, E., Garcia, A.: Phys. Lett. B {\bf 464}, 25(1999)\\
Banerjee, R., Majhi, B.R.: J. High Energy Phys. {\bf 06}, 095(2008)\\
Banerjee, R., Modak, S.K.: J. High Energy Phys. {\bf 0905}, 063(2009)\\
Bekenstein, J.D.: Nuovo Cimento Lett. {\bf 4}, 737(1972)\\
Bronnikov, K.A.: Phys. Rev. Lett. {\bf 85}, 4641(2000)\\
Bytsenko, A.A., et al.: Phys. Lett. B {\bf 443}, 121(1998a)\\
Bytsenko, A.A., et al.: Phys. Rev. D {\bf 57}, 4917(1998b)\\
Bytsenko, A.A., et al.: Phys. Rev. D {\bf 64}, 105024(2001)\\
Chen, Y.-X., et al.: Europhys. Lett. {\bf 95}, 10008(2011)\\
Cognola, G., et al.: Phys. Rev. D {\bf 52}, 4548(1995)\\
Elizalde, E., et al.: Phys. Rev. D {\bf 59}, 061501(1999)\\
Gibbons, G.W., Hawking, S.W.: Phys. Rev. D {\bf 15}, 2752(1977)\\
Hawking, S.W.: Nature {\bf 248}, 30(1974)\\
Hartle, J.B., Hawking, S.W.: Phys. Rev. D {\bf 13}, 2188(1976)\\
Iyer, V., Wald, R.M.: Phys. Rev. D {\bf 50}, 846(1994)\\
Jacobson, T., et al.: Phys. Rev. D {\bf 49}, 6587(1994)\\
Jacobson, T., et al.: Phys. Rev. D {\bf 52}, 3518(1995)\\
Jamil, M., Darabi, F.: Int. J. Theor. Phys. {\bf 50}, 2460(2011)\\
Kothawala, D., et al.: Phys. Lett. B {\bf 652}, 338(2007)\\
Larra\~{n}aga, A.: Commun. Theor. Phys. {\bf 55}, 72(2011a)\\
Larra\~{n}aga, A.: Pramana J. Phys. {\bf 76}, 553(2011b)\\
Majhi, B.R.: Phys. Rev. D {\bf 79}, 044005(2009)\\
Majhi, B.R., Samanta, S.: Annals Phys. {\bf 325}, 2410(2010)\\
Matyjasek, J.: Phys. Rev. D {\bf 76}, 084003(2007)\\
Matyjasek, J.: Acta Physica Polonica B {\bf 39}, 3(2008)\\
Modak, S.K.: Phys. Lett. B {\bf 671}, 167(2009)\\
Nickalls, R.W.D.: The Mathematical Gazette. {\bf 77}, 354(1993)\\
Nojiri, S., Odintsov, S.D.: Phys. Rev. D {\bf 59}, 044003(1999a)\\
Nojiri, S., Odintsov, S.D.: Phys. Lett. B {\bf 463}, 57(1999b)\\
Nojiri, S., Odintsov, S.D.: Int. J. Mod. Phys. A {\bf 15}, 989(2000)\\
Nojiri, S., Odintsov, S.D.: Int. J. Mod. Phys. A {\bf 16}, 1015(2001)\\
Parikh, M.K.: Gen. Relativ. Gravit. {\bf 36}, 2419(2004) [Int. J. Mod.
Phys.\\ D {\bf 13}, 2351(2004)]\\
Parikh, M.K., Wilczek, F.: Phys. Rev. Lett. {\bf 85}, 5042(2000)\\
Sharif, M., Javed, W.: J. Korean Phys. Soc. {\bf 57}, 217(2010)\\
Sharif, M., Javed, W.: Canadian J. Phys. (to appear, 2011)\\
Srinivasan, K., Padmanabhan, T.: Phys. Rev. D {\bf 60}, 024007(1999)\\
Visser, M.: Phys. Rev. D {\bf 46}, 2445(1992)\\
Visser, M.: Phys. Rev. D {\bf 48}, 583(1993a)\\
Visser, M.: Phys. Rev. D {\bf 48}, 5697(1993b)\\
Wald, R.M.: Phys. Rev. D {\bf 48}, 3427(1993)\\
Whitt, B.: Phys. Rev. D {\bf 32}, 379(1985)\\
Zhu, T., et al.: Corrected Entropy of
High Dimensional Black Holes.\\ arXiv:0906.4194v2 (2009a)\\
Zhu, T., et al.: JCAP {\bf0908}, 010(2009b)\\

\end{document}